\documentclass[journal,12pt,onecolumn,draftclsnofoot,]{IEEEtran}
\usepackage{float}
\usepackage{lipsum}
\usepackage[utf8]{inputenc}
\usepackage{color}
\usepackage{graphicx}
\usepackage{amssymb}
\usepackage{multicol}
\usepackage{amsmath}
\usepackage{amsthm}
\usepackage[cmintegrals]{newtxmath}
\usepackage[normalem]{ulem}

\usepackage{mathtools}

\usepackage[letterpaper, left=0.68in, right=0.68in, bottom=1.024in, top=0.7in]{geometry}

\newcommand\blfootnote[1]{%
  \begingroup
  \renewcommand\thefootnote{}\footnote{#1}%
  \addtocounter{footnote}{-1}%
  \endgroup
}

\definecolor{dblue}{rgb}{0,0,0.8}

\definecolor{red}{rgb}{1,0,0}

\begin{document}
\title{On the Effect of Mutual Coupling in One-Bit\\
Spatial Sigma-Delta Massive MIMO Systems\\
}
\author{Hessam Pirzadeh$^{*}$, Gonzalo Seco-Granados$^{\dagger}$, A. Lee Swindlehurst$^{*}$, and Josef A. Nossek$^{\S}$\\
$^{*}$Center for Pervasive Commun. \& Computing, UC Irvine, USA.
Email: \{hpirzade, swindle\}@uci.edu\\
$^{\dagger}$Dept.~of Telecom. \& Systems Eng., Univ. Aut\`{o}noma de Barcelona, Spain. Email: gonzalo.seco@uab.cat\\
$^{\S}$Dept. of Teleinformatics Eng., Federal Univ. of Ceara, Brazil. Email: josef.a.nossek@tum.de}

\maketitle
\begin{abstract}
 The one-bit spatial Sigma-Delta concept has recently been proposed as an approach for achieving low distortion and low power consumption for massive multi-input multi-output systems. The approach exploits users located in known angular sectors or spatial oversampling to shape the quantization noise away from desired directions of arrival. While reducing the antenna spacing alleviates the adverse impact of quantization noise, it can potentially deteriorate the performance of the massive array due to excessive  mutual coupling. In this paper, we analyze the impact of mutual coupling on the uplink spectral efficiency of a spatial one-bit Sigma-Delta massive MIMO architecture, and compare the resulting performance degradation to standard one-bit quantization as well as the ideal case with infinite precision. Our simulations show that the one-bit Sigma-Delta array is particularly advantageous in space-constrained scenarios, can still provide significant gains even in the presence of mutual coupling when the antennas are closely spaced.
%\vspace{-1mm}
\end{abstract}

%\begin{IEEEkeywords}
%Massive MIMO, analog-to-digital converter, spectral efficiency.
%\end{IEEEkeywords}
%\blfootnote{This work was supported in part by the National Science Foundation under Grant ECCS-1547155 and Grant CCF-1703635 and in part
%by a Hans Fischer Senior Fellowship from the Technische Universit\"at M\"unchen
%Institute for Advanced Study. The work of J. A. Nossek has been performed in the framework of the Horizon
%2020 project ONE5G (ICT-760809) receiving funds from the European Union.}
\blfootnote{This work was supported by the U.S. National Science Foundation under Grants CCF-1703635 and ECCS-1824565. 
%The work of G. Seco-Granados has been partially funded by the Spanish Ministry of Science and Universities Grant PRX18/00638. The work of J. A. Nossek has been performed in the framework of the Horizon
%2020 project ONE5G (ICT-760809) receiving funds from the European Union.
}
%\vspace{-7mm}

\section{Introduction}\label{sec:Introduction}

%\vspace{-1mm}
Power consumption is a key concern for next-generation wireless networks. Deploying power efficient base stations (BSs) while satisfying high-data-rate demands is of crucial importance. Massive multiple-input multiple-output (MIMO) architectures are considered to be an important component of next-generation systems to meet the aforementioned objectives, but the high power consumption required by large arrays employing high-resolution analog-to-digital converters (ADCs) poses some practical problems. Hence, the use of low-resolution ADCs with reduced power consumption has gained attention in recent years \cite{Fan}-\cite{LiTSML17}. 
%{\bf (add more references to fill in some of the available space)}

Recently, a spatial Sigma-Delta ($\Sigma\Delta$) architecture has been proposed to compensate for the performance loss due to the use of one-bit quantization in massive MIMO systems \cite{baracspatial}-\cite{RaoSP19}. In this architecture, either the users of interest are assumed to lie within some known angular sector (e.g., as in a sectorized wireless cell), or the array is assumed to be spatially oversampled (antennas spaced less than one-half wavelength $\lambda$ apart), so that a spatial analog of the classical $\Sigma\Delta$ approach can be used to shape the quantization noise to angles away from the users' desired directions of arrival (DoAs). Unlike temporal oversampling, there is a limit to the amount of spatial oversampling that can be achieved, due to the physical dimensions of the antennas. In addition, the impact of mutual coupling may become significant as the antenna spacing decreases \cite{Masorous,Biswas}. While \cite{Hessam_JSAC} has shown that the low-complexity one-bit $\Sigma\Delta$ architecture can achieve performance approaching that of systems with high-resolution ADCs, this prior work did not consider the mutual coupling effect. 

In this paper, we investigate the impact of mutual coupling on the performance of the spatial $\Sigma\Delta$ approach assuming a uniform linear array (ULA) of antennas. Our results show that the one-bit Sigma-Delta array is particularly advantageous in space-constrained scenarios, and can still provide significant gains even in the presence of mutual coupling when the antennas are closely spaced. For very small antenna spacings, the noise shaping gain is offset by the loss due to mutual coupling, and the performance remains relatively constant; this is in contrast to a standard high-resolution ADC architecture without $\Sigma\Delta$, where the performance degrades monotonically as the antennas move closer together.

\emph{Notation}: We use boldface letters to denote vectors, and capitals to denote matrices. The $(i,j)$-th element of matrix $\boldsymbol{A}$ and the $i$-th element of vector $\boldsymbol{a}$ are denoted by $[\boldsymbol{A}]_{ij}$ and $a_i$, respectively. The symbols $(.)^*$, $(.)^T$, and $(.)^H$ represent conjugate, transpose, and conjugate transpose, respectively. A circularly-symmetric complex Gaussian (CSCG) random vector with zero mean and covariance matrix $\mathit{\boldsymbol{R}}_{\boldsymbol{v}}$ is denoted by $\boldsymbol{v}\sim\mathcal{CN}(\boldsymbol{0},\boldsymbol{\boldsymbol{R}_{\boldsymbol{v}}})$. $\mathrm{Ci}(x)\triangleq\gamma + \mathrm{log}(x)+\int_{0}^{x}{\frac{\mathrm{cos}(t) - 1}{t}dt}$, and $\mathrm{Si}(x)\triangleq\int_{0}^{x}{\frac{\mathrm{sin}(t)}{t}dt}$ denote cosine and sine integrals where $\gamma$ is the Euler-Mascheroni constant.
$\mathbb{E}[.]$, $\mathfrak{R}\{.\}$ and $\mathfrak{I}\{.\}$ represent the expectation operator, the real part and imaginary part of a complex value, respectively. We use $\mathrm{diag}\left(\boldsymbol{x}\right)$ as the diagonal matrix formed from the elements of vector $\boldsymbol{x}$.

%\vspace{-3mm}

\section{System Model}\label{sec:SYSTEM MODEL}
\subsection{Channel Model and Mutual Coupling}
%\vspace{-1mm} 
Consider the uplink of a single-cell multi-user MIMO system consisting of $K$ single-antenna users that send their signals simultaneously to a BS equipped with a uniform linear array (ULA) with ${M}$ antennas. The $M\times 1$ signal received at the BS from the $K$ users is given by
\begin{equation}\label{channel model}
  {\boldsymbol{x}}=\boldsymbol{G}\boldsymbol{P}^{\frac{1}{2}}
  \boldsymbol{{s}} + {\boldsymbol{n}},
\end{equation}
where $\boldsymbol{G}=\left[\boldsymbol{g}_1,\cdots,\boldsymbol{g}_K\right]\in\mathbb{C}^{M\times K}$ is the channel matrix between the users and the BS, and $\boldsymbol{P}$ is a diagonal matrix whose $k$th diagonal element, $p_k$, represents the transmitted power of the $k$-th  user. The scalar symbols transmitted by the users are collected in the vector $\boldsymbol{s}\in\mathbb{C}^{K\times 1}$, where $\mathbb{E}\left\{\boldsymbol{s}\boldsymbol{s}^H\right\}=\boldsymbol{I}_K$ and the symbols are assumed to be independently drawn from a CSCG codebook.
%$x_k$ is the symbol transmitted from the $k$th user and is drawn from $\mathit{\boldsymbol{x}}\in\mathbb{C}^{K\times1}$ which satisfies $\mathbb{E}\{\mathit{\boldsymbol{x}}\mathit{\boldsymbol{x}}^{H}\}=\boldsymbol{I}_K$, 
The term ${\boldsymbol{n}}\sim\mathcal{CN}\left(\boldsymbol{0},\boldsymbol{R}_{\boldsymbol{n}}\right)$ denotes additive CSCG receiver noise at the BS. 

For the $k$th user, the channel vector is modeled as
\begin{equation}\label{channel model_phy}
\boldsymbol{g}_k=\sqrt{\frac{\beta_k}{L}}\boldsymbol{T}\boldsymbol{A}_k\boldsymbol{h}_k,	
\end{equation}
%\begin{equation}\label{channel model_phy}
%\boldsymbol{G}=\boldsymbol{T}\boldsymbol{A}\boldsymbol{H}\boldsymbol{D}^{\frac{1}{2}},	
%\end{equation}
where $\beta_k$ models geometric attenuation and shadow fading, the columns of the $M\times L$ matrix $\boldsymbol{A}_k$ are steering vectors corresponding to signal arrivals from different DoAs, and $\boldsymbol{h}_k$ represents fast-fading channel coefficients that are assumed to be independently and identically distributed as $\mathcal{CN}\left({0},1\right)$ random variables. The matrix $\boldsymbol{T}$ models the mutual coupling:
\begin{equation}\label{mc_mat}
	\boldsymbol{T} = \left(\boldsymbol{I} + \frac{1}{R}\boldsymbol{Z} \right)^{-1}
\end{equation}
where $R$ denotes the low-noise amplifier (LNA) input impedance. Assuming thin half-wavelength dipoles, the elements of $\boldsymbol{Z}$ can be characterized as \cite{Schelkunoff}
\begin{equation*}\label{mutual_coupling}
	[\boldsymbol{Z}]_{ij} = 30\biggl(2\mathrm{Ci}(2\pi d_{ij}) - \mathrm{Ci}(\xi_{ij}+\pi) - \mathrm{Ci}(\xi_{ij} - \pi)\\
	+ j\left(-2\mathrm{Si}(2\pi d_{ij}) + \mathrm{Si}(\xi_{ij}+\pi) + \mathrm{Si}(\xi_{ij} - \pi)\right)
	\biggr),~i\neq j
\end{equation*}
\begin{equation}
[\boldsymbol{Z}]_{ii} = 30\bigl(\gamma +\mathrm{log}(2\pi)-\mathrm{Ci}(2\pi)+j\mathrm{Si}(2\pi)\bigr),	
\end{equation}
where $d_{ij}$ is the distance between antennas $i$ and $j$ normalized by the wavelength, and $\xi_{ij}=\pi\sqrt{1+4d_{ij}^{2}}$.
 
Since we will be focusing on situations where the signals of interest arrive within a certain angular sector, we assume a physical channel model in which the angular domain for each user is described by $L$ fixed DoAs, as in \cite{Ngo_PHY}. Hence, $\boldsymbol{A}_k$ has columns defined by the steering vectors
\begin{equation}\label{steering vector}
\boldsymbol{a}\left(\theta_{k\ell}\right)=
%\frac{1}{\sqrt{L}}%\begin{bmatrix}
%        e^{-jf_1(\varphi_p)},e^{-jf_2(\varphi_p)},\cdots,e^{-jf_M(\varphi_p)}
%\end{bmatrix}^T,	
[1,e^{-j{2\pi d_{12}}\mathrm{sin}\left(\theta_{k\ell}\right)},\cdots, e^{-j{2\pi d_{1M}}\mathrm{sin}\left(\theta_{k\ell}\right)}]^T,
\end{equation}
for DoAs $\theta_{k\ell}$.
The channels for each user are further distinguished by independent fast fading coefficients, which we collect in the matrix $\boldsymbol{H}=\left[\boldsymbol{h}_1,\cdots,\boldsymbol{h}_K\right]\in\mathbb{C}^{L\times K}$. 
%In the remainder of the paper, we assume that $\boldsymbol{A}$ and $\boldsymbol{D}$ are priori known at the BS \cite{Zhang,Ngo_PHY}. 

The presence of mutual coupling will not only affect the channel, but will also in general produce colored noise at the receiver. For the mutual coupling model described above, the covariance $\boldsymbol{R}_{\boldsymbol{n}}$ of the additive noise can be derived as \cite{Ivrlac}
%{\bf (need a citation)}
\begin{equation}\label{R_n}
\boldsymbol{R}_{\boldsymbol{n}}=\\
\boldsymbol{T}\left(\sigma_i^2\left(\boldsymbol{Z}\boldsymbol{Z}^H + R_N^2\boldsymbol{I} - 2R_N\mathfrak{R}(\rho^*\boldsymbol{Z})\right)+4\mathrm{k}TB\mathfrak{R}(\boldsymbol{Z})\right)\boldsymbol{T}^H	,
\end{equation}
with $\mathbb{E}\{\boldsymbol{i}_N\boldsymbol{i}_N^H\}=\sigma_i^2\boldsymbol{I}$, $\mathbb{E}\{\boldsymbol{u}_N\boldsymbol{u}_N^H\}=\sigma_u^2\boldsymbol{I}$, $R_N=\frac{\sigma_u}{\sigma_i}$, $\mathbb{E}\{\boldsymbol{u}_N\boldsymbol{i}_N^H\}=\frac{\rho}{\sigma_u\sigma_i}\boldsymbol{I}$, where $\boldsymbol{i}_N$ and $\boldsymbol{u}_N$ denote the equivalent noise current and voltage of the low noise amplifier (LNA), and $\mathrm{k}$, $T$, and $B$ represent the Boltzman constant, environment temperature, and bandwidth, respectively.  

\subsection{$\Sigma\Delta$ Quantization}
In a standard implementation involving one-bit quantization, each antenna element at the BS is connected to a one-bit ADC. In such systems, the received baseband signal at the $m$th antenna becomes
\begin{equation}\label{quantizer}
    y_m = \mathcal{Q}_m\left(x_m\right) \; ,
\end{equation}
where $\mathcal{Q}_m\left(.\right)$ denotes the one-bit quantization operation applied separately to the real and imaginary parts as
\begin{equation}
\mathcal{Q}_m\left(x_m\right)=\alpha_{m,r}\mathrm{sign}\left(\mathfrak{R}\left(x_m\right)\right)+j\alpha_{m,i}\mathrm{sign}\left(\mathfrak{I}\left(x_m\right)\right) \; .
\end{equation}
{The output voltage levels of the one-bit quantizers are represented by $\alpha_{m,r}$ and $\alpha_{m,i}$}. While the output level is irrelevant for standard one-bit quantization, for $\Sigma\Delta$ quantization the selection of adequate output levels is of critical importance, as discussed in \cite{Hessam_JSAC}. {Furthermore,} we allow these levels to be a function of the antenna index $m$, although once the values of $\alpha_{m,r}$ and $\alpha_{m,i}$ are chosen, they remain fixed and independent of the user scenario or channel realization. After quantization, the received baseband signal at the BS is given by
%\color{black}
%(\GS{Internal comment: I have modified this, introducing $\alpha_{m,r}$ and $\alpha_{m,i}$, instead of defining $\alpha_m$ as a complex number. The advantage is that later we can use $\alpha_m$ both for the real and imaginary outputs.})
%$\mathcal{Q}_m\left(.\right)=\mathfrak{Re}\left(\alpha_m\right)\mathrm{sign}\left(\mathfrak{Re}\left(x_m\right)\right)+j\mathfrak{Im}\left(\alpha_m\right)\mathrm{sign}\left(\mathfrak{Im}\left(x_m\right)\right)$, with $\alpha_m$ as the output quantization level. Therefore, the received baseband signal at the BS is
\begin{equation}\label{vector-quantizer}
    \boldsymbol{y}=\mathcal{Q}\left(\boldsymbol{x}\right) = 
    \begin{bmatrix}
           \mathcal{Q}_1\left(x_1\right),
           \mathcal{Q}_2\left(x_2\right),
           \cdots,
           \mathcal{Q}_M\left(x_M\right)
    \end{bmatrix}^T.
\end{equation}
%\begin{equation}\label{vector-quantizer}
%    \boldsymbol{y}=\mathcal{Q}\left(\boldsymbol{x}\right) = 
%    \begin{bmatrix}
%           \mathcal{Q}_1\left(x_1\right) \\
%           \mathcal{Q}_2\left(x_2\right) \\
%           \vdots \\
%           \mathcal{Q}_M\left(x_M\right)
%    \end{bmatrix}.
%\end{equation}
%\vspace{-1mm}

%\vspace{-2mm}
By appropriately designing the output voltages of the ADCs (see \cite{Hessam_JSAC} for details), the received baseband signal at the BS after $\Sigma\Delta$ quantization becomes
\begin{equation}\label{shaped-equation}
    \boldsymbol{y} = \boldsymbol{x} + \boldsymbol{U}^{-1}\boldsymbol{q} \; ,
\end{equation}
in which 
\begin{equation}\label{U-matrix}
\boldsymbol{U}=
  \left[ {\begin{array}{ccccc}
   1 & & & & \\
   e^{-j\phi} & 1 & & & \\
   \vdots & \ddots & \ddots & & \\
   e^{-j\left(M-1\right)\phi} & \cdots & e^{-j\phi} & 1 & \\
  \end{array} } \right] \; ,
\end{equation}
where $\phi$ denotes the center angle of the sector with low quantization noise, and $\boldsymbol{q}$ represents the effective quantization noise. Following the same reasoning as in \cite{Hessam_JSAC}, the covariance matrix of the quantization noise can be approximated as
\begin{equation}
    \boldsymbol{R}_{\boldsymbol{q}}\simeq\mathrm{diag}\left(\boldsymbol{p}_{\boldsymbol{q}}\right),
\end{equation}
where 
\begin{eqnarray}
    \boldsymbol{p}_{\boldsymbol{q}} & = & \left(\frac{\pi}{2}\zeta-1\right)\boldsymbol{\Pi}\boldsymbol{p}_{\boldsymbol{x}} \\
\boldsymbol{p}_{\boldsymbol{x}} & = & 
\begin{bmatrix}
        \mathbb{E}\left[|x_1|^2\right], \mathbb{E}\left[|x_2|^2\right],\cdots,\mathbb{E}\left[|x_M|^2\right]
\end{bmatrix}^T 
%\left[
%{\begin{array}{c}
%\matrix \mathbb{E}\left[|\chi_1|^2\right] %\cr \mathbb{E}\left[|\chi_2|^2\right] \cr %\vdots \cr %\mathbb{E}\left[|\chi_M|^2\right]
%\end{array}}\right],
\end{eqnarray}
\begin{equation*}
\bf{\Pi}=\\
    \left[
    \begin{array}{cccccc}
    \begin{matrix} 
    1 & & & & & \bf{0} \cr
    \left(\frac{\pi}{2}\zeta-1\right) & 1 & & & &  \cr
    \vdots & \ddots & 1 & & & \cr \left(\frac{\pi}{2}\zeta-1\right)^{m} & \ddots& \ddots & \ddots & & \cr \vdots& \ddots & \ddots& \ddots & \ddots & \cr \left(\frac{\pi}{2}\zeta-1\right)^{M-1} &\cdots & \left(\frac{\pi}{2}\zeta-1\right)^{m}& \cdots&\left(\frac{\pi}{2}\zeta-1\right) & 1
    \end{matrix}
    \end{array} \right]
\end{equation*}
and $\zeta=1.13$ is a correction factor.
In the following sections, we investigate the spectral efficiency of the system described above and study the impact of antenna spacing.

\section{Spectral Efficiency}
Due to the complicated structure of the mutual coupling matrix in (\ref{mc_mat}) and the quantization noise shaping matrix $\boldsymbol{U}^{-1}$, a closed-form expression for the spectral efficiency (SE), if it exists, would likely not provide significant insight into its behavior with respect to antenna spacing, nor would it provide a tool for the purpose of optimization. Hence, in the next section we numerically evaluate the SE of the system. 
 
The received $\Sigma\Delta$-quantized signal, ${\boldsymbol{y}}$, at the BS is
\begin{equation}
\begin{aligned}
    {\boldsymbol{y}} = \mathcal{Q}\left({\boldsymbol{x}} \right) = {\boldsymbol{G}}\boldsymbol{P}^{\frac{1}{2}} {\boldsymbol{s}} + {\boldsymbol{n}} + {\boldsymbol{ U}}^{-1} {\boldsymbol{q}} \; .
    \end{aligned}
    \label{Eq22}
\end{equation}
The total effective noise $\boldsymbol{\eta}={\boldsymbol{n}} + {\boldsymbol{ U}}^{-1} {\boldsymbol{q}}$ has covariance matrix $\boldsymbol{R}_{\boldsymbol{\eta}}=\boldsymbol{R}_{\boldsymbol{n}}+
{\boldsymbol{ U}}^{-1} \boldsymbol{R}_{\boldsymbol{q}} {\boldsymbol{ U}}^{-H}$.
We assume the BS employs a linear receiver $\boldsymbol{W}$, and we will consider the case of maximum ratio combining (MRC) and zero-forcing (ZF). For MRC, we do not account for the fact that $\boldsymbol{R}_{\boldsymbol{\eta}}$ is spatially colored, since pre-whitening $\boldsymbol{G}$ destroys the approximate orthogonality of the array response and increases the inter-user interference. Thus, for MRC we set ${\boldsymbol{W}} = {\boldsymbol{G}}$. However, knowledge of $\boldsymbol{R}_{\boldsymbol{\eta}}$ can be exploited by the ZF receiver, and thus we assume the pre-whitened solution ${\boldsymbol{W}} = \boldsymbol{R}_{\boldsymbol{\eta}}^{-1} {\boldsymbol{G}}({\boldsymbol{G}}^H \boldsymbol{R}_{\boldsymbol{\eta}}^{-1} {\boldsymbol{G}})^{-1}$.

For either receiver, the detected symbol vector is
\begin{equation}
 \hat{\boldsymbol{s}} = {\boldsymbol{W}}^H {\boldsymbol{y}} =  {\boldsymbol{W}}^H{\boldsymbol{ G}}\boldsymbol{P}^{\frac{1}{2}} {\boldsymbol{s}} +  {\boldsymbol{W}}^H{\boldsymbol{n}} + {\boldsymbol{ W}}^H{\boldsymbol{U}}^{-1} {\boldsymbol{q}}.
    \label{Eq23}
\end{equation}
The $k$-th detected symbol can be written as
\begin{equation}\label{detected}
\begin{aligned}
    \hat{s}_k = &\sqrt{p_k}{\boldsymbol{w}}_k^H {\boldsymbol{g}}_k s_k + 
     \sqrt{p_k} {\boldsymbol{w}}_k^H \sum_{i \neq k} {\boldsymbol{g}}_i s_i +  {\boldsymbol{w}}_k^H {\boldsymbol{n}} +  {\boldsymbol{w}}_k^H {\boldsymbol{U}}^{-1} {\boldsymbol{q}},
    \end{aligned}
\end{equation}
where $\boldsymbol{w}_k$ is the $k$th column of $\boldsymbol{W}$. We assume the BS treats ${{\boldsymbol{w}}}_{k}^H{{\boldsymbol{g}}}_{k}$ as the desired signal and the other terms of (\ref{detected}) as worst-case Gaussian noise when decoding the signal. Consequently, a lower bound for the ergodic achievable SE at the $k$th user can be written as \cite{Matthaiou}
{\begin{equation}\label{SE_bound}
 \mathcal{S}_k =\\
 \begin{aligned}
 \mathbb{E} \left[{\rm log}_2\left( 1 + \frac{p_k\left\lvert{\boldsymbol{w}}_k^H {\boldsymbol{g}}_k \right\rvert^2}{ \sum_{i \neq k}p_i  |{\boldsymbol{w}}_k^H {\boldsymbol{g}}_i|^2  +  \left\lvert {\boldsymbol{w}}_k^H {\boldsymbol{n}} \right\rvert^2 + \left\lvert {\boldsymbol{w}}_k^H {\boldsymbol{U}}^{-1} {\boldsymbol{q}} \right\rvert^2  } \right)\right] \; .
\end{aligned}
\end{equation}}
%\begin{figure*}[!t]
%\begin{equation}\label{approx_theo}
%\mathcal{S}_k\approx\mathrm{log}_2\left(1+\frac{p_k\beta_k\left(\left|\mathrm{Tr}\left[\boldsymbol{\Sigma}_{kk}\right]\right| ^2 + \mathrm{Tr}\left[\boldsymbol{\Sigma}_{kk}^{2}\right]\right)}{\sum_{i=1,i\ne k}^{K}{p_i\beta_i\mathrm{Tr}\left[\boldsymbol{\Sigma}_{kk}\boldsymbol{\Sigma}_{ii}\right]}+\sigma_n^2\mathrm{Tr}\left[\boldsymbol{\Sigma}_{kk}\right]+\mathrm{Tr}\left[\boldsymbol{\Sigma}_{kk}\boldsymbol{U}^{-1}\boldsymbol{R}_{\boldsymbol{q}}\boldsymbol{U}^{-H}\right]}\right)	
%\end{equation}
%\hrulefill
%\vspace*{4pt}
%\end{figure*}
%\begin{figure*}[!t]
%\begin{equation}\label{ZF_SE}
%\mathcal{S}_k=\mathbb{E}\left[\mathrm{log}_2\left(1+\frac{p_k}{\sigma_n^2\left[\left(\boldsymbol{G}^H\boldsymbol{G}\right)^{-1}\right]_{kk}+\left[\left(\boldsymbol{G}^H\boldsymbol{G}\right)^{-1}\boldsymbol{G}^H\boldsymbol{U}^{-1}\boldsymbol{R}_{\boldsymbol{q}}\boldsymbol{U}^{-H}\boldsymbol{G}\left(\boldsymbol{G}^{H}\boldsymbol{G}\right)^{-1}\right]_{kk}}\right)\right]
%\end{equation}
%\hrulefill
%\vspace*{4pt}
%\end{figure*}

\section{Numerical Results}\label{sec:numerical results}
In this section, we numerically evaluate the SE performance of the $\Sigma\Delta$ massive MIMO system for various scenarios. We assume static-aware power control in the network \cite{Emil Bj} so that $p_k=p_0/\beta_k$. In all of the cases considered, unless otherwise noted, we assume $K=10$ users and equally spaced antennas with normalized spacing $d$. The DoAs for each user are drawn uniformly from the interval $[\theta_{0}-\delta,\theta_{0}+\delta]$, and the center angle of the $\Sigma\Delta$ array is steered towards $\phi=2\pi{d}\mathrm{sin}\left(\theta_0\right)$. 

To highlight the impact of mutual coupling, we will compare the performance when mutual coupling is included to that when it is hypothetically absent. Simulating the case without mutual coupling amounts to setting $\boldsymbol{Z}=R\boldsymbol{I}$ in \eqref{mc_mat} and~(\ref{R_n}), which leads to $\boldsymbol{T}=\frac{1}{2}\boldsymbol{I}$ and $\boldsymbol{R}_{\boldsymbol{n}}=\sigma_n^2\boldsymbol{I}$, with the noise power given by
\begin{equation*}
\sigma_n^2=\frac{1}{4}\left[\sigma_i^2\left(R^2+R_N^2-2R_NR\mathfrak{R}(\rho)\right)+4kTBR\right].
\end{equation*}
The factor of $1/2$ in $\boldsymbol{T}$ results from the fact that $\boldsymbol{x}$ in~\eqref{channel model} is the voltage on a load matched to the antenna impedance. This voltage is half of the antenna open-circuit voltage, and given that the load represents the input to the LNA, it is the signal available for further processing. Thus, the per-antenna and per-user  reference signal-to-noise-ratio (SNR) in the absence of mutual coupling is given by
\begin{equation}
  \mathrm{SNR}\triangleq \frac{1}{4} \frac{p_0}{\sigma_n^2}. 
\end{equation}
The circuit parameters used in~(\ref{mc_mat}) and~(\ref{R_n}) are defined as $\sigma_i^2=2kTB/R$, and $\sigma_u^2=2kTBR$, leading to $R_N=R$ where $R=50~\Omega$, $T=290~\mathrm{K}$, $\rho=0$, and $B=20~\mathrm{MHz}$. This leads to a value of $\sigma_n^2 = 2kTBR$, where the factor of~2 appears because we are accounting for noise in both the antennas and the LNAs. We further assume CSCG symbols and $10^4$ Monte Carlo trials for the simulations.

%To highlight the impact of mutual coupling, we will compare the performance when mutual coupling is included to that when it is hypothetically absent. Simulating the case without mutual coupling amounts to setting $\boldsymbol{Z}=R\boldsymbol{I}$ in eq.~\eqref{mc_mat}, which leads to $\boldsymbol{T}=\frac{1}{2}\boldsymbol{I}$. The factor of $1/2$ results due to the assumption of a load impedance matched to that of the antenna, which means that the power of the signal at the receiver is $1/2$ of that at the antenna input. We will define the reference signal-to-noise-ratio (SNR) as that which is present at the receiver in the case with no mutual coupling, which is thus given by
%\begin{equation}
%  \mathrm{SNR}\triangleq \frac{1}{2} \frac{p_0}{\sigma_n^2}, 
%\end{equation}
%where the noise power 
%\begin{equation*}
%\sigma_n^2=\frac{1}{4}\left[\sigma_i^2\left(R^2+R_N^2-2R_NR\mathfrak{R}(\rho)\right)+4kTBR\right]
%\end{equation*}
%results from substituting $\boldsymbol{Z}=R\boldsymbol{I}$ and $\boldsymbol{T}=\frac{1}{2}\boldsymbol{I}$ into~(\ref{R_n}), which leads to $\boldsymbol{R}_{\boldsymbol{n}}=\sigma_n^2\boldsymbol{I}$. The circuit parameters used in~(\ref{mc_mat}) and~(\ref{R_n}) are defined as $\sigma_i^2=2kTB/R$, and $\sigma_u^2=2kTBR$, leading to $R_N=R$ where $R=50~\Omega$, $T=290~\mathrm{K}$, $\rho=0$, and $B=20~\mathrm{MHz}$. This leads to a value of $\sigma_n^2 = 2kTBR$, where the factor of~2 appears because we are accounting for noise in both the antennas and the LNAs. We further assume CSCG symbols and $10^4$ Monte Carlo trials for the simulations.

\begin{figure}
\centering
\includegraphics[width=1\textwidth]
{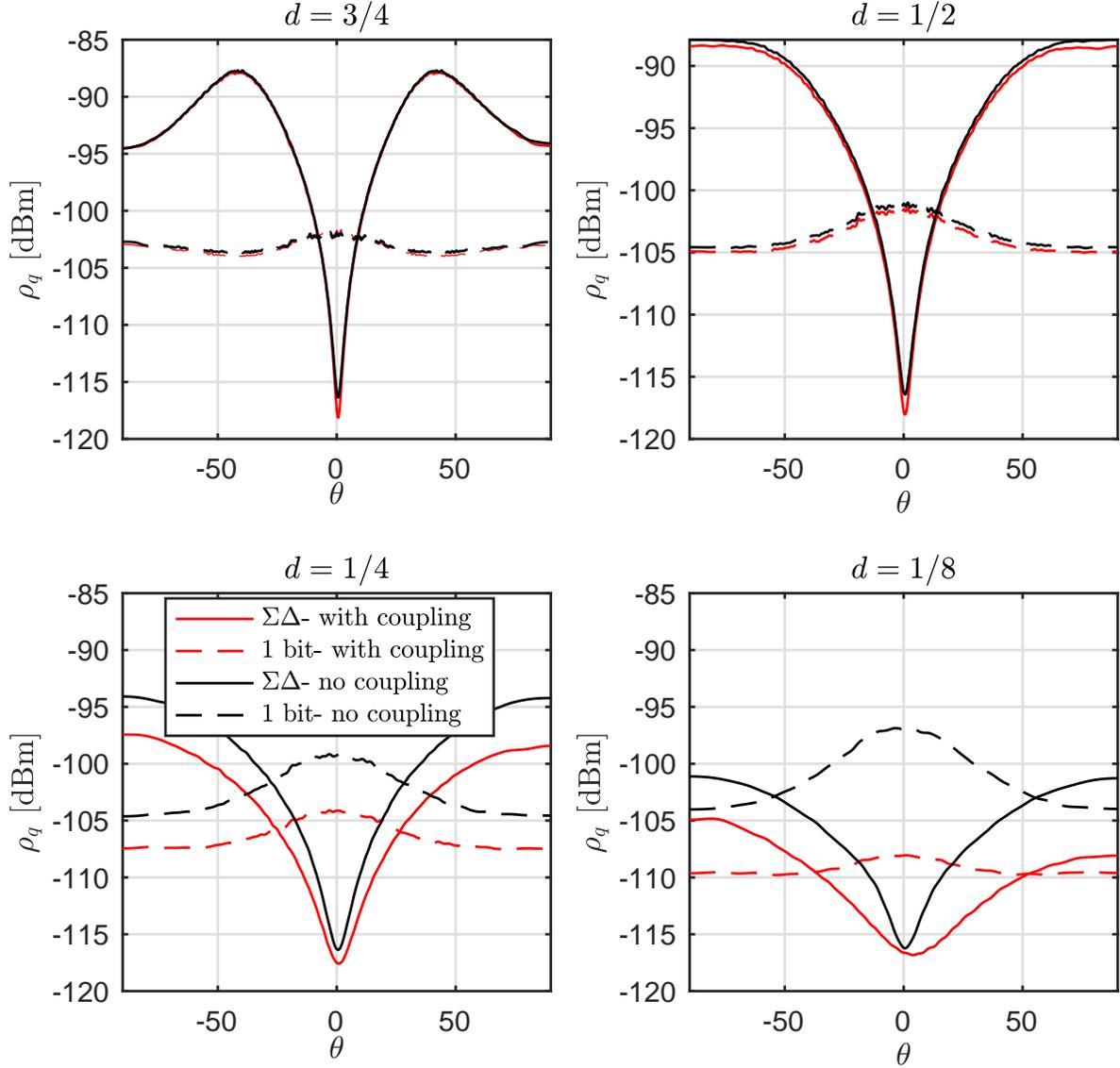}
\caption{Quantization noise power density for a system with $\theta_0=0^{\circ}$, $2\delta=40^{\circ}$, $\mathrm{SNR}=~0~\mathrm{dB}$, $M=100$, $L=15$.}
\label{fig1}
\end{figure}

In Fig. \ref{fig1}, we investigate the impact of the mutual coupling matrix, $\boldsymbol{T}$, on the spatial spectrum of the quantization noise when $\theta_0=0$. 
To do so, we define the quantization noise power density as
\begin{equation}\label{noise-density}
   \rho_q\left(\theta\right)\triangleq\frac{1}{c(\theta)}\mathbb{E}\left[\left|\boldsymbol {a}\left(\theta\right)^H\boldsymbol{T}^H\boldsymbol{U}^{-1}\boldsymbol{q}\right|^2\right],
\end{equation}
where $c(\theta)=\left\|\boldsymbol{T}\boldsymbol{a}\left(\theta\right)\right\|^2$ is a normalizing factor and $\theta\in\left[-90^{\circ},90^{\circ}\right]$ denotes the DoA. We see that the noise shaping characteristic of the $\Sigma\Delta$ array is not significantly affected by the mutual coupling, except for the case of $d=\lambda/8$, where we see a small shift in the quantization noise spectrum.

\begin{figure}
\centering
\includegraphics[width=1\textwidth]
{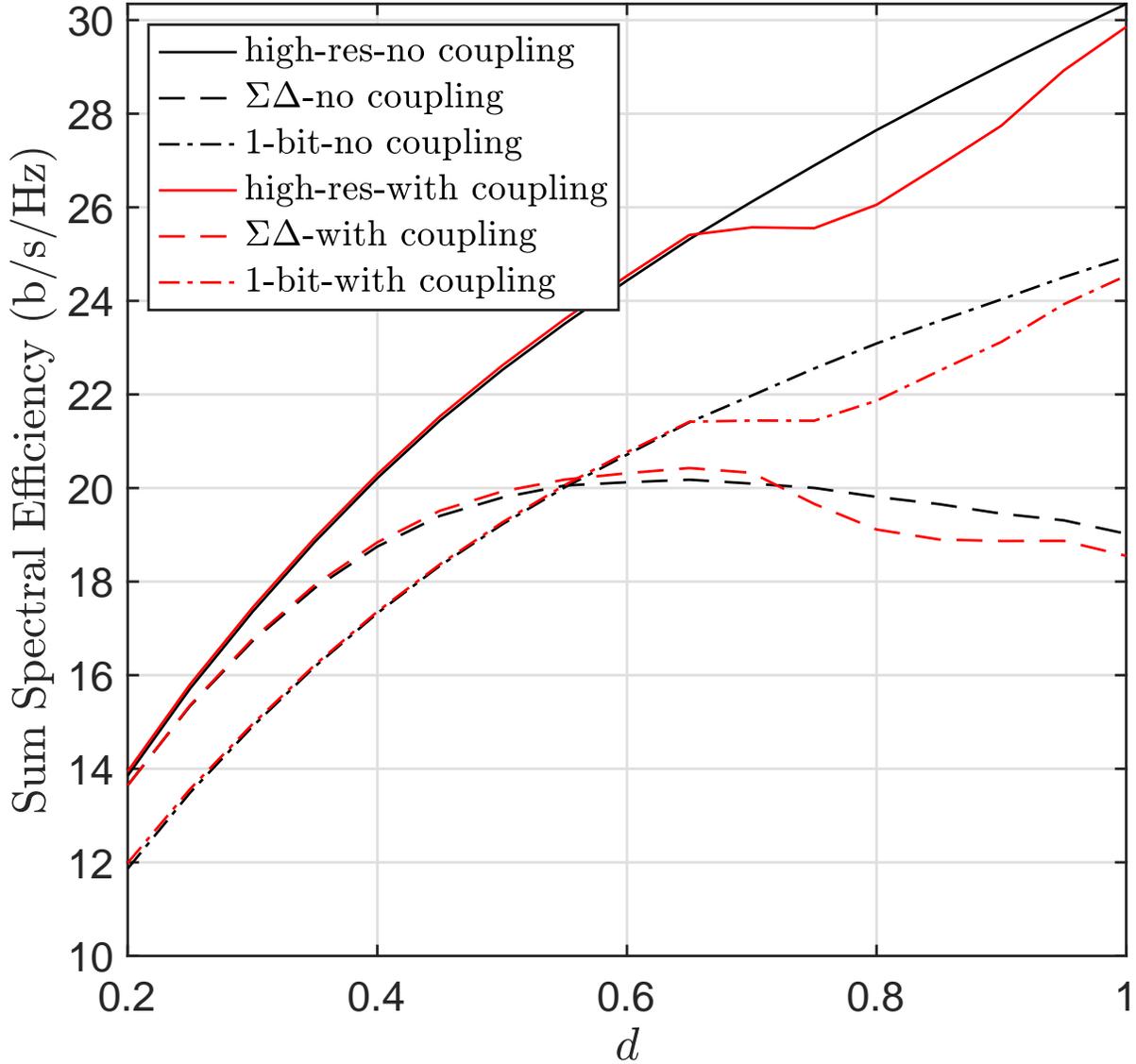}
\caption{SE versus antenna spacing for a system with an MRC receiver and $\theta_0=-10^{\circ}$, angular sector $2\delta=40^{\circ}$, $\mathrm{SNR}=10~\mathrm{dB}$, $M=100$, $L=15$.}
\label{fig2}
\end{figure}
In Fig.~\ref{fig2},
%\GS{\textbf{(As before, and also for the other figures, adding the lines obtained without coupling would be very useful in order to understand the effects of coupling with sigma-delta quantizers.)}}
we show the effect of antenna spacing on the SE of a system with an MRC receiver. 
%{\bf (I don't think that is what Fig. 2 is showing).} 
We see that, when there is no constraint on the size of the array, better performance for the standard one-bit architecture can be achieved by moving the antennas farther apart. 
%\GS{(Is the figure showing this? I observe that for the $\Sigma\Delta$ the SE decreases if the antennas are more than $0.6\lambda$ apart)}
We see that the standard one-bit architecture outperforms the $\Sigma\Delta$ array when $d > \lambda/2$, due to the fact that increasing the antenna spacing increases the quantization noise power for the $\Sigma\Delta$ architecture across the DoA sector of interest, as observed in Fig.~\ref{fig1}. Furthermore, we see that the SE for the $\Sigma\Delta$ architecture is not monotonic and $d=\lambda/2$ provides the best performance, which corresponds to no oversampling. The optimal value of $d$ for the $\Sigma\Delta$ array will of course decrease if the sector of user DoAs was widened.

\begin{figure}
\centering
\includegraphics[width=1\textwidth]
{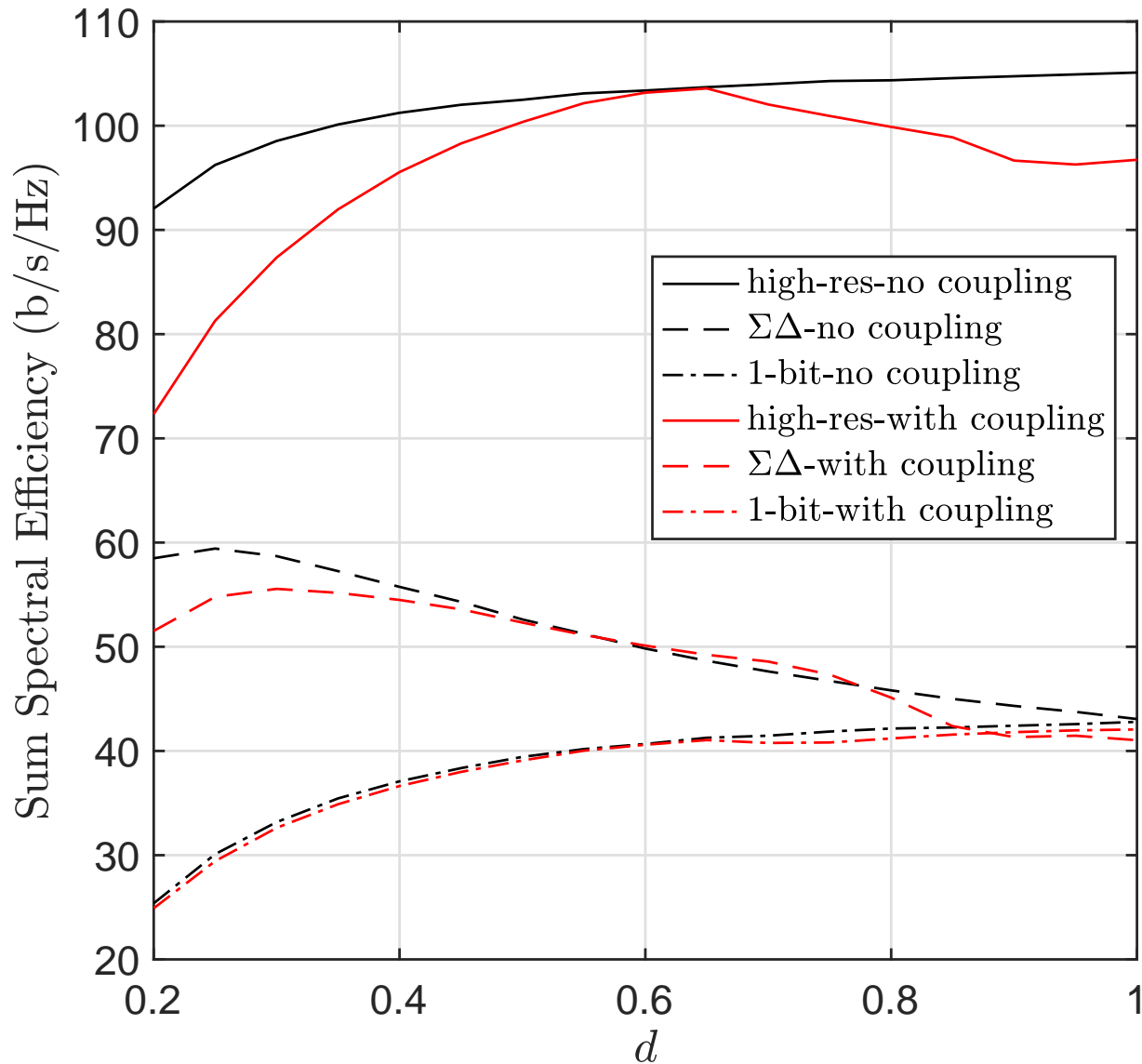}
\caption{SE versus antenna spacing for a system with ZF receiver and $\theta_0=-10^{\circ}$, angular sector $2\delta=40^{\circ}$, $\mathrm{SNR}=~10~\mathrm{dB}$, $M=100$, $L=15$.}
\label{fig3}
\end{figure}

The SE results for the ZF receiver are shown in Fig.~\ref{fig3}. We again observe the  degradation of the $\Sigma\Delta$ performance as $d$ increases, but in this case there is a more significant gain relative to standard one-bit quantization for smaller antenna spacings, and the optimal antenna spacing for the $\Sigma\Delta$ array is reduced to approximately $d=\lambda/3$.

%{\bf (since you have room, show this with a simulation rather than in words; i.e., show the optimal spacing for MRC and ZF as a function of SNR).} 
\begin{figure}
\centering
\includegraphics[width=1\textwidth]
{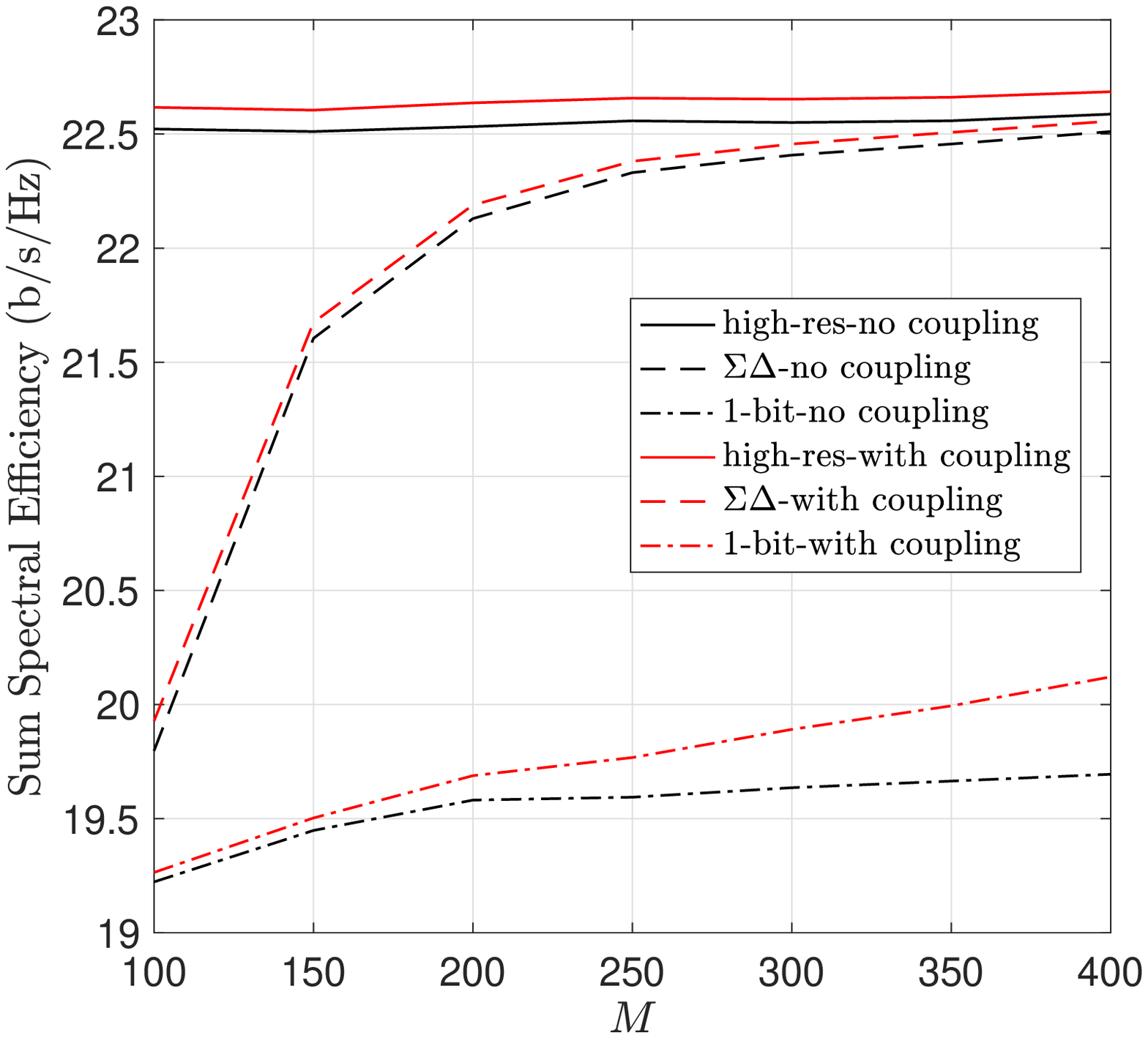}
\caption{SE versus number of BS antennas for a system with MRC receiver and $\theta_0=-10^{\circ}$, angular sector $2\delta=40^{\circ}$, $\mathrm{SNR}=10~\mathrm{dB}$, $d_0=50$, $L=15$.}
\label{fig4}
\end{figure}

\begin{figure}
\centering
\includegraphics[width=1\textwidth]
{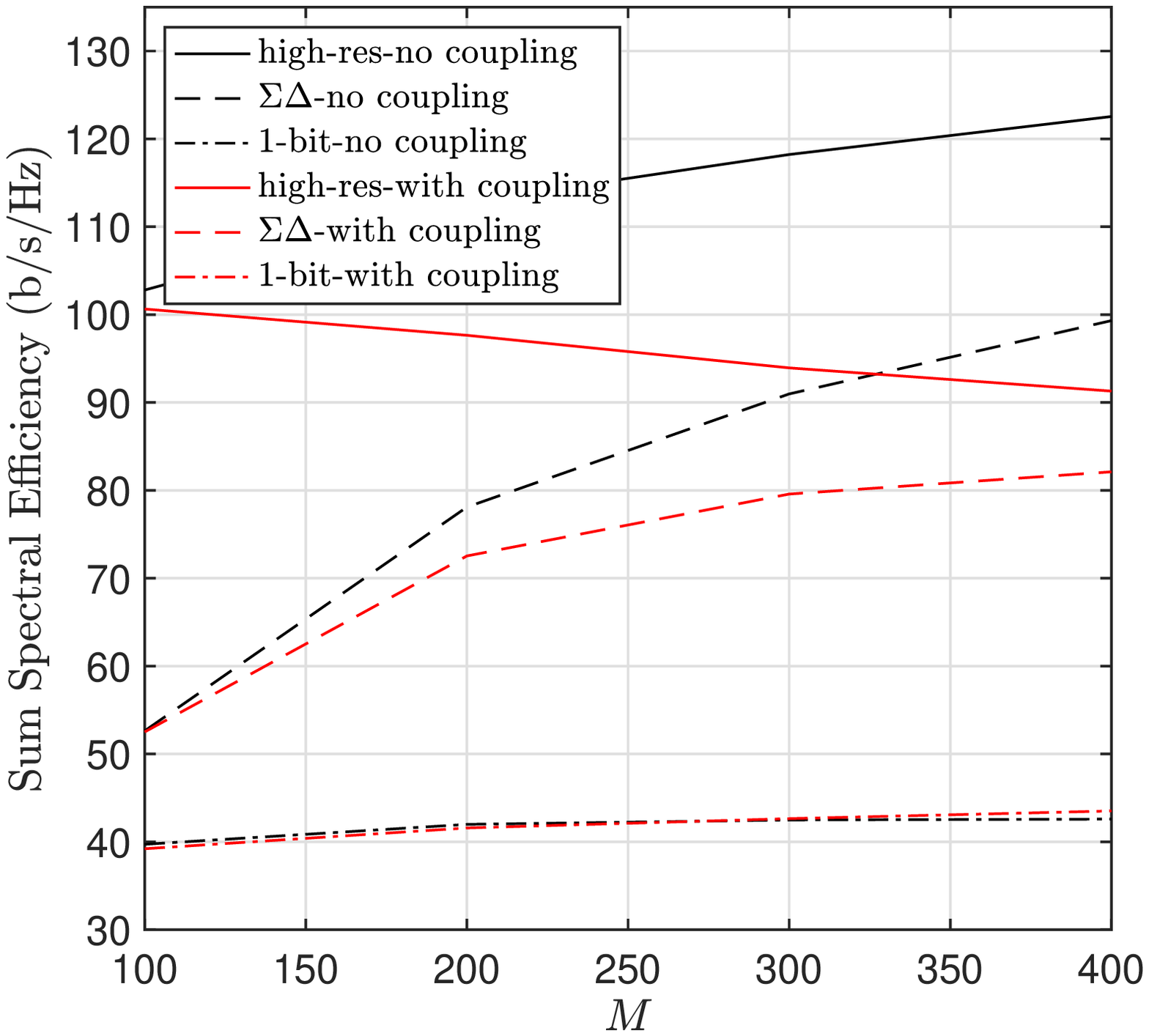}
\caption{SE for a system with ZF receiver and $\theta_0=-10^{\circ}$, angular sector $2\delta=40^{\circ}$, $\mathrm{SNR}=10~\mathrm{dB}$, $d_0=50$, $L=15$.}
\label{fig5}
\end{figure}

While the standard one-bit architecture can outperform the $\Sigma\Delta$ approach when there is no constraint on the dimension of the array (large $d$), Figs.~\ref{fig4} and~\ref{fig5} demonstrate that the $\Sigma\Delta$ array provides a better result in space-constrained scenarios. For these simulations, we consider a case in which the antenna array has a limited aperture of $d_0=50$ and we increase the number of antennas from $M=100$ to $M=400$, which corresponds to a decrease in antenna spacing from $d=1/2$ to $d=1/8$. For the case of an MRC receiver in Fig.~\ref{fig4}, the  $\Sigma\Delta$ architecture achieves a spectral efficiency nearly equal to that of an array with full-resolution ADCs when $M \ge 250$. For the case of ZF, $\Sigma\Delta$ provides a dramatic gain in SE over standard one-bit quantization.

\begin{figure}
\centering
\includegraphics[width=1\textwidth]
{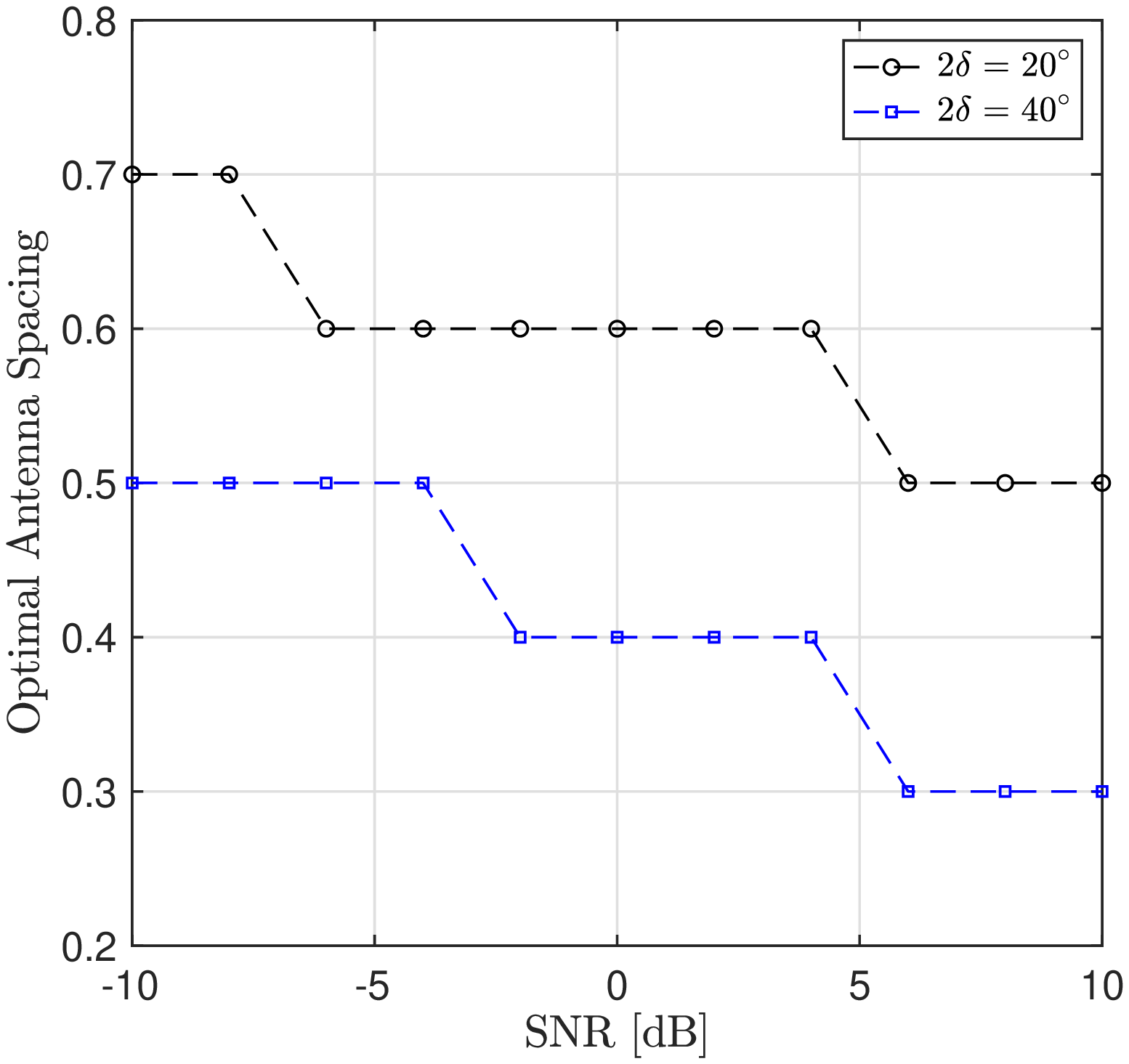}
\caption{Optimal antenna spacing versus SNR for a system with ZF receiver and $\theta_0=-10^{\circ}$, $M=100$, $L=15$.}
\label{fig6}
\end{figure}

In Fig.~\ref{fig6}, the optimal antenna spacing for the $\Sigma\Delta$ architecture with a ZF receiver is shown for different SNRs, where we have quantized $d$ to the nearest value of $\lambda/10$. It can be seen that the optimal spacing is dependent on the SNR and DoA region width, $\delta$. The optimal antenna spacing decreases as SNR increases, and also as the size of the DoA sector of interest increases. We expect this phenomenon since for wider DoA sectors, a wider noise shaping characteristic is required to achieve the best performance. The same general conclusion holds true for the case with the MRC receiver. 

\section{Conclusion}
We {have} studied the effect of mutual coupling on the performance of one-bit $\Sigma\Delta$ massive MIMO systems. It was shown that the $\Sigma\Delta$ architecture is most suitable for array deployments with an aperture size constraint. While the performance of standard one-bit quantization saturates as the number of antennas increases in a constrained-aperture array, the performance of the $\Sigma\Delta$ architecture tends to approach that of a system with high-resolution ADCs. This is due to the noise-shaping gain achieved by the $\Sigma\Delta$ architecture when the users are sectorized or the array is oversampled in space. It is worthwhile to note that the inevitable power loss due to mutual coupling can to some extent be alleviated using, for example, a matching network. This is a subject of future investigation.

\color{black}

\end{document}